\RequirePackage{etex}
\documentclass{aa}
\usepackage{graphicx}
\usepackage{multirow}
\usepackage{txfonts}
\usepackage{lipsum}
\usepackage{lscape}             
\usepackage{placeins}           
 \usepackage{hyperref} 

 \usepackage{multirow}
\usepackage{array}
\usepackage{amsmath}

\begin{document}

  \title{Influence of CO versus CH$_4$ on organic haze formation in atmospheres of diverse terrestrial exoplanets}


   \author{Sai Wang\inst{1}
        \and Zhengbo Yang\inst{1}
        \and Haixin Li\inst{1}
        \and Chao He\inst{1}\fnmsep\thanks{Corresponding author:chaohe23@ustc.edu.cn}
        \and Yingjian Wang\inst{1}
        \and Xiao’ou Luo\inst{1}
        \and Yu Liu\inst{1}
        \and Sarah M. Hörst\inst{2,3}
        \and Sarah E. Moran\inst{4,3}
        \and Véronique Vuitton\inst{5}
        \and Laurène Flandinet\inst{5}
        \and Patricia McGuiggan\inst{6}
        }

\institute{
School of Earth and Space Sciences, University of Science and Technology of China, Hefei 230026, China
\and
Department of Earth and Planetary Sciences, Johns Hopkins University, Baltimore, MD, USA
\and
Space Telescope Science Institute, Baltimore, MD, USA
\and
NHFP Sagan Fellow, NASA Goddard Space Flight Center, Greenbelt, MD 20771, USA
\and
Univ. Grenoble Alpes, CNRS, IPAG, 38000 Grenoble, France
\and
Department of Materials Science and Engineering, Johns Hopkins University, Baltimore, MD, USA
}
    

 
  \abstract
   {Terrestrial exoplanets are expected to host secondary, high metallicity atmospheres derived from outgassing of volatiles like N$_2$, CO$_2$, H$_2$O, CH$_4$, and CO. Photochemical organic hazes are likely to form in such environments, significantly impacting both atmospheric observation and planetary habitability.}
   {This study aims to investigate haze formation across representative terrestrial exoplanet atmospheres and assess how CH$_4$ versus CO as the primary carbon source differentially affects haze production rates, particle properties, and chemical complexity.} 
   {We conducted six laboratory simulations by exposing the initial gas mixture (a few mbar) to glow discharge at 300 K. Each simulated atmosphere  comprised 75\% N$_2$, CO$_2$, or H$_2$O, 10\% of each of the other two gases, and 5\% CH$_4$ or CO. We analyzed the gas-phase products using a residual gas analyzer. For solid products, we measured production rates and particle density, determined particle size distributions via atomic force microscopy, identified functional groups using Fourier-transform infrared spectroscopy, and characterized molecular composition with very high-resolution mass spectrometry. }
   {Experiments using CH$_4$ produced a wider diversity of gas-phase species and substantially higher haze yields compared to the corresponding CO-based experiments. CO-derived haze particles exhibited a restricted size range (10--80 nm), whereas CH$_4$-derived hazes formed denser material with complex functional group signatures and thousands of unique molecular formulas. The pattern of the identified molecular formulas indicates molecular growth pathways linked to detected gaseous precursors such as HCN, CH$_2$O, and C$_2$H$_4$.}
   {The atmospheric redox state critically controls haze formation in simulated terrestrial exoplanet atmospheres. CH$_4$ is significantly more effective than CO in initiating organic growth, leading to higher haze production rates and greater chemical complexity. These results provide crucial constraints for exoplanet atmospheric modeling and spectral interpretation, and further support the possibility that reducing atmospheres may facilitate prebiotic organic chemistry relevant to the emergence of life.}


   \maketitle

\section{Introduction}

Exploring planetary habitability and searching for extraterrestrial life are widely recognized as important scientific frontiers. To date, over 6000 exoplanets have been confirmed, including more than 200 Earth-sized planets. Some of these terrestrial planets, orbiting within the habitable zones of their host stars, may possess conditions suitable for life, such as Proxima Centauri b \citep{2016_Anglada-Escude_Natur}, Kepler-186 f \citep{2014_Quintana_Sci}, K2-72 e \citep{2017_Dressing_AJ}, TRAPPIST-1 e, f, and g \citep{2016_Gillon_Natur, 2017_Gillon_Natur}, TOI-700 d and TOI-700 e \citep{2020_Gilbert_AJ, 2023_Gilbert_ApJL}. The discovery of these terrestrial planets provides opportunities to seek possible signs of life. Given the great distances to these exoplanets, we can only probe the potential biosignatures by characterizing their atmospheric properties via telescope observation in the foreseeable future.Current and future telescope surveys are expected to find many more such potentially life-supporting candidates, which will be central targets for atmospheric characterization with both ground- and space-based observatories, including JWST, ELTs, Ariel, and HWO. 

Observations have shown that clouds and/or hazes are present in atmospheres of many sub-Neptunes and gas giant exoplanets, such as in K2-18b \citep{2025_Jaziri_A&A, 2025_Liu_arXiv}, GJ 1214b \citep{2014_Kreidberg_Natur, 2023_Gao_ApJ, 2023_Kempton_Natur}, HD 97658b \citep{2014_Knutson_ApJ}, GJ 436b \citep{2014_Knutson_Nature}, GJ 3470b \citep{2015_Dragomir_ApJ}, WASP 39b \citep{2023_Alderson_Natur}, and HATS 8b \citep{2020_May_AJ}. These particles can absorb, scatter, and reflect radiation across different wavelengths, impacting atmospheric spectra of exoplanets as well as their temperature profile and habitability \citep{2019_Adams_ApJ, 2019_Kawashima_ApJL, 2021_Gao_JGRE, 2017_Arney_ApJ, 2017_Zhang_Natur}. Cloud and haze particles are also expected in terrestrial exoplanet atmospheres, based on the understanding of terrestrial planets in our solar system. Although there are a number of equilibrium cloud species that may form at lower planetary temperatures, observational, theoretical, and experimental studies have revealed that photochemical organic hazes tend to form in small, cool atmospheres with high metallicity \citep{2013_Marley_CCTP, 2014_Kreidberg_Natur, 2017_Morley_ApJ, 2018_He_AJ, 2018_Horst_NatAs, 2020_Gao_ApJ, 2025_Ohno_ApJL}. These hazes may contain prebiotic molecules, providing organic materials for origin of life on distant worlds \citep{2012_Horst_AsBio, 2020_Moran_PSJ}. Therefore, it is vital to understand the efficiency of photochemical haze formation in atmospheres of terrestrial exoplanets and their compositional and optical properties to enable accurate atmospheric characterization with current and future telescopes. Laboratory production of exoplanet hazes can help elucidate the plausible formation mechanisms, compositions, and optical characteristics of haze particles, thereby providing valuable insights to guide and interpret future atmospheric observations.

Previous laboratory investigations have explored haze formation in simulated N$_2$-dominated terrestrial (exo)planet atmospheres. For instance, \citet{2004_Trainer_AsBio} investigated how CO$_2$ concentration affects haze composition in CH$_4$/CO$_2$/N$_2$ gas mixtures. Gavilan et al. (\citeyear{2017_Gavilan_APJL}, \citeyear{2018_Gavilan_ApJ}) extended these studies, further examining how different CO$_2$/CH$_4$ ratios influence haze composition and optical properties. More recently, \citet{2026_Drant_A&A} conducted a cross-laboratory study to measure the refractive indices of hazes produced from gas mixtures with different N$_2$/CH$_4$ and CH$_4$/CO ratios, demonstrating that both the experimental setup and the degree of nitrogen incorporation profoundly influence the optical properties and chemical signatures of hazes.

While these studies have laid a crucial foundation for understanding how different carbon sources influence haze formation in N$_2$-dominated atmospheres, terrestrial exoplanets are expected to exhibit a much broader diversity of atmospheric compositions. Therefore, this study aims to provide a systematic investigation into the roles of CO and CH$_4$ as primary carbon sources in haze formation across diverse terrestrial exoplanet atmospheres. Specifically, we conducted experiments with initial gas mixtures representative of H$_2$O-rich, N$_2$-rich, and CO$_2$-rich atmospheres. During the experiments, we monitored the gas composition using a residual gas analyzer (RGA). After each experiment, solid products were collected to determine yield and density, and the size distributions of solid particles were measured using atomic force microscopy (AFM). The chemical composition of the haze particles was then measured by a vacuum Fourier-transform infrared spectrometer (FTIR) and a very high-resolution mass spectrometer (VHRMS). By analyzing both gas and solid-phase products, we investigate haze formation processes and highlight the correlations linking gas-phase precursors to haze growth and composition in terrestrial exoplanets. 
\section{Methods}

\subsection{Initial gas compositions and experimental setup}
The atmospheres of the terrestrial exoplanets are likely to be diverse based on theoretical expectations of the stochastic nature of planet formation and volatile delivery in putative protoplanetary disks \citep{2015_Leconte_Sci}. Observations and modeling suggest enhanced elevated metallicity in these diverse atmospheres \citep{2010_Schaefer_Icar, 2014_Forget_RSPTA, 2018_deWit_NatAs} because primary atmospheres (formed by accretion of gaseous matter from the accretion disc) are probably lost due to a combination of surface temperature, mass of the atoms and escape velocity of the planet. Thus, these planets are likely to have secondary atmospheres that originated by the outgassing of internal volatiles, such as CO, CO$_2$, CH$_4$, H$_2$O, and N$_2$ \citep{2014_Gaillard_E&PSL, 2023_Liggins_JGRE}. Due to the complexity and quantities of unknown parameters of terrestrial exoplanets, we focused on three of the most probable atmospheric compositional scenarios to investigate haze formation in these planets. The main difference between the three scenarios was the background gas (75\%, similar to Earth’s N$_2$ level): either H$_2$O-rich, N$_2$-rich or CO$_2$-rich. After establishing the background gas, the two remaining species were added in equal quantities at 10\% levels of each. 

Then, either CH$_4$ or CO, selected depending on the oxidizing/reducing power of the atmosphere, was added at 5\% level as a carbon source for organic hazes to ensure that they participate in photochemistry at an observable level in a short experimental period. CH$_4$ is often considered as a necessary carbon source for organic haze formation. However, previous studies \citep{2019_He_ESC, 2018_Horst_NatAs, 2019_Fleury_APJ} have demonstrated that CH$_4$ is not required for organic haze formation and CO can act as an alternative carbon source. CO is likely to be a major carbon-bearing species produced by outgassing on rocky planets \citep{2014_Gaillard_E&PSL, 2024_Tian_ApJ, 2023_Liggins_JGRE}, making it a highly relevant carbon source for terrestrial exoplanets. Although volcanic outgassing is less likely to supply abundant CH$_4$ to the atmosphere \citep{2020_Wogan_PSJ}, thermochemical equilibrium and atmospheric evolution models suggest that CH$_4$ can become the dominant carbon-bearing species under relatively reducing and cooler conditions, and thus remains an important carbon source to consider \citep{2013_Moses_ApJ, 2023_Liggins_JGRE, 2025_Bower_AJ}. 

\begin{figure}[ht!]
   \centering
   \includegraphics[width=\hsize]{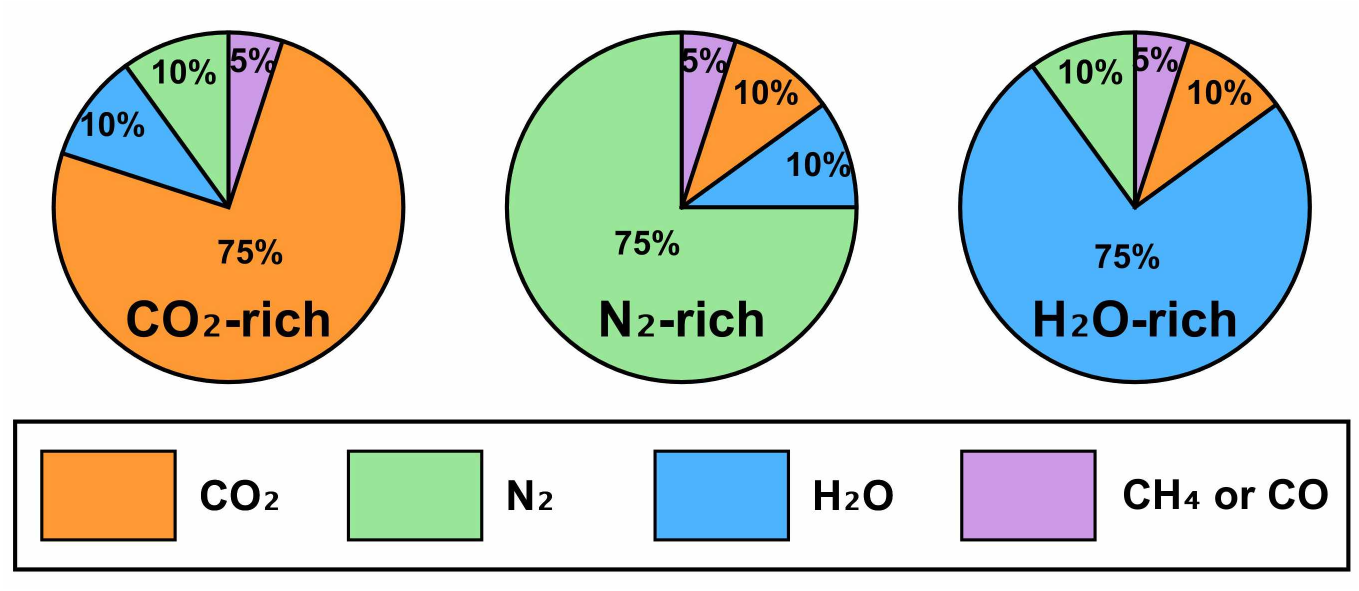}
      \caption{Initial gas mixing ratios for the six experiments. Each mixture consists of a dominant species (CO$_2$, N$_2$, or H$_2$O; 75\%) and the remaining two background gases (10\% each), combined with 5\% of either CH$_4$ or CO.}
         \label{fig1}
   \end{figure}

Here, we ran experiments separately with either CH$_4$ or CO as carbon sources to investigate their role in organic haze formation. For convenience, each experiment was labeled by pairing the dominant atmospheric component with the carbon source: N$_2$-rich/CO and N$_2$-rich/CH$_4$ for N$_2$-dominated runs, CO$_2$-rich/CO and CO$_2$-rich/CH$_4$ for CO$_2$-dominated runs, and H$_2$O-rich/CO and H$_2$O-rich/CH$_4$ for H$_2$O-dominated runs. This resulted in a total of six experiments, with the specific compositions shown in Fig.~\ref{fig1}. Although there are other possible atmospheric compositions, here we focused on these six atmospheric composition scenarios as a starting point to explore haze formation in a range of terrestrial exoplanets. We ran experiments at 300 K, representing the equilibrium temperature of habitable terrestrial exoplanets. We performed all experiments using the PHAZER setup with the established experimental procedure described in our previous studies \citep{2017_He_ApJL, 2018_He_AJ}. The initial gas mixture was flowed into a stainless steel chamber at 10 standard cubic centimeters per minute (sccm) and exposed to an AC glow discharge at a pressure of a few mbar. The gas phase products flowed out of the chamber and resulting gas phase products were either vented to a fume hood or measured by an RGA.

Solid phase products were deposited on the chamber walls and on pre-positioned substrates. Mica and Calcium Fluoride (CaF$_2$) substrates were placed in the chamber before each experiment to facilitate subsequent analysis (Section 2.3). Following the completion of each experimental run, we stopped the discharge, the gas flow, and the temperature control, but kept the chamber under vacuum for 48 hours to remove the volatile components before sample collection. After each experiment, the chamber was transferred to a dry (<0.1 ppm H$_2$O), oxygen free (<0.1 ppm O$_2$) N$_2$ glove box (Inert Technology Inc., I-lab 2GB) for sample collection. For experiments with high haze production rates, all deposits from both the chamber walls and substrates were collected. In contrast, for experiments with low haze production rates where solid particles were insufficient for collection and weighing, only the substrates with deposited films were retrieved for further characterization.

\subsection{Gas-phase product analysis}

During the experiments, we continuously monitored the composition of gases flowing out of the chamber over a mass-to-charge (m/z) range of 1-100 using an RGA (a quadrupole mass spectrometer, Stanford Research Systems, Inc.) with a 70 eV electron ionization (EI) source. Prior to each experiment, the RGA was used to detect the background signal of the reaction system and verify the purity and mixing ratio of the reaction gases. After the plasma discharge stabilized, we performed at least two acquisitions, each an average of 300 scans, to confirm the stability of the system and ensure the consistency of the results.

The unit mass resolution of the measured mass spectra leads to an overlap of different gaseous products, making it difficult to unambiguously identify the species formed. Therefore, we deconvolved the mass spectra using an established Monte Carlo method \citep{2020_Gautier_RCMS, 2020_Serigano_DPS, 2022_Serigano_JGRE}, with reference to the National Institute of Standards and Technology (NIST) mass spectral database. A tolerance of ±40\% in fragment ion intensities was applied to compensate for instrumental response differences. Similar approaches have been successfully used in previous studies to analyze RGA data \citep{2020_Bourgalais_NatSR, 2022_He_ESC, 2025_Wang_ApJ}. 

We focused our analysis on the m/z < 47 range, as peak intensities at higher mass approached the background noise level ($\sim 1.0 \times 10^{-10}$ arbitrary units). Within this range, the standard library included 49 candidate gaseous species containing C, H, O, and N elements that could potentially contribute to the mass spectra. We used all possible species to fit the entire spectrum simultaneously, thereby identifying the set of species that provided the best overall match.  To enable comparison across the six experiments, the total intensity of each mass spectrum was normalized. After millions of iterations, the 5000 best fits (with the smallest residuals) were selected, the top 20\% of which were averaged to represent the probable gaseous product composition. It should be noted that the reported values represent highly approximate relative abundances rather than absolute concentrations, given the unknown uncertainties associated with EI fragmentation patterns. These results are primarily intended for qualitative purposes rather than precise quantification and are aimed at identifying the most probable species. 

\subsection{Solid-phase product analysis}

For the experiments that produced weighable solid particles, the haze density was determined by measuring their mass and volume using a high-precision analytical balance (Sartorius) and a gas pycnometer (AccuPyc II 340, Micrometrics), respectively. The production rate was also calculated by dividing the total collected mass by the reaction time (mg/hour), which represents a lower limit because it is impossible to completely remove all particles from the chamber. For the low production rate cases, we analyzed the particles deposited on the mica substrates using  an atomic force microscope (Bruker Dimension 3100) with supersharp tips (SHR 150 probes, Budget Sensors, tip radius $< 1\,\mathrm{nm}$, cone angle $< 20^\circ$). Mica was selected because its atomically flat surface facilitates high resolution AFM imaging of haze films. From the AFM images of the film samples, we characterized the particle size distributions and particle counts, allowing us to calculate the total particle volume by assuming spherical particles uniformly distributed over the inner chamber surfaces. Then we estimated the particle mass and production rate with an assumed particle density ($\rho = 1.50\,\mathrm{g\,cm^{-3}}$, close to similar haze samples, \citep{2017_He_ApJL, 2025_Wang_ApJ}). Note that the estimated haze production rates also represent lower limits, as AFM-based particle counts only capture the top layer of particles on the film.

Decades of research on planetary haze have shown that these materials are extremely chemically complex. To identify complex functional groups of haze particles, we measured the transmittance spectra of the solid products by employing a vacuum Fourier-transform infrared spectrometer (FTIR, Bruker VERTEX 70V). For high production rate experiments, we used KBr pellet methods for the spectral measurements following previous procedures \citep{2024_He_NatAs}. Collected solid products were ground and mixed with KBr powder to produce a homogeneous sample/KBr mixture, which was then pressed into pellets for FTIR analysis. For the low production rate cases, we measured the transmittance of the haze-deposited CaF$_2$ films. We acquired the spectra in the mid infrared range of $3600$--$1000~\mathrm{cm}^{-1}$ ($2.8$--$10~\mu\mathrm{m}$) with a resolution of $1\,\mathrm{cm^{-1}}$ using a KBr beamsplitter and a DLaTGS detector. In this wavelength range, both KBr and CaF$_2$ are optically transparent while most organic molecules exhibit characteristic absorption bands, allowing the identification of various functional groups in different solid samples. 

Further, we obtained the very high-resolution mass spectra using an LTQ-Orbitrap XL (Thermo Scientific) with mass accuracy $\pm 1~\mathrm{ppm}$ and over a mass range of $150$--$800~\mathrm{amu}$. The collected solid powders were dissolved in methanol (CH$_3$OH, $1~\mathrm{mg\,mL^{-1}}$) and centrifuged at $10000~\mathrm{rpm}$ for $10~\mathrm{min}$, while the films sample were immersed in $1~\mathrm{mL}$ of CH$_3$OH for $24~\mathrm{hr}$ before collecting the resulting CH$_3$OH-sample mixture. The soluble fraction of the samples was measured with the Orbitrap in both negative and positive ionization modes, which helps reduce measurement bias \citep{2020_Moran_PSJ, 2021_Vuitton_PSJ, 2022_Moran_JGRE}. ``Blank'' solutions were measured before each sample as a control to monitor possible background contamination. The acquired data exhibit thousands of peaks with very high mass resolving power ($m/\Delta m \geq 1.0 \times 10^5$). Following the established protocol described in our previous studies \citep{2025_Wang_ApJ, 2025_Yang_PSJ}, we obtained accurate molecular formulas in each sample and calculation of double bond equivalents (DBE). This approach provided precise molecular-level characterization of the soluble organic fraction, enabling identification of specific molecular species and their elemental compositions.

\section{Results and Discussion}

\subsection{Composition of Gas-phase Products}

\begin{figure*}[t!]
    \centering
    \resizebox{16cm}{!}
    {\includegraphics{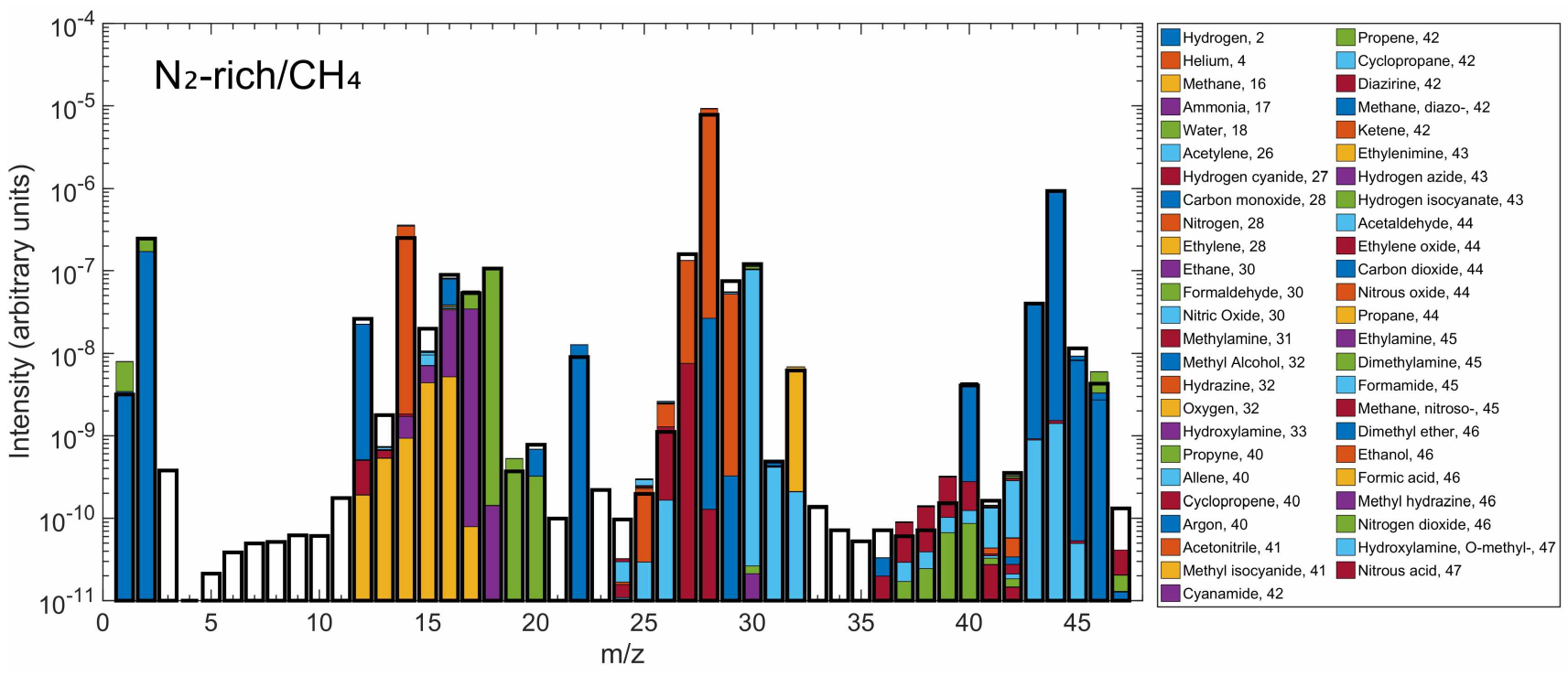}}
    \caption{Example of deconvolution results for the N$_2$-rich/CH$_4$ experiment. Black outline bars show the normalized mass spectrum, while colored sections indicate the model-fitted contribution of each identified species to its corresponding mass channel. Low intensity channels, shown as open black bars, represent background noise or masses not associated with stable products in the database. Species in the legend are ordered with their molecular mass. The O$_2$ (m/z 32) and Ar (m/z 40) peaks are attributed to the RGA background, as they are present without additional gas injection and do not increase in the plasma-on spectrum relative to the plasma-off spectrum.}
    \label{fig2}
\end{figure*}

Fig.~\ref{fig2} shows an example of mass spectral deconvolution for the N$_2$-rich/CH$_4$ plasma discharge experiment. The deconvolved mass spectrum reveals a variety of gaseous species, including both the initial mixture and newly formed compounds. Deconvolution was performed using an iterative Monte Carlo fitting algorithm, which was applied consistently across all six experiments \citep{2025_Wang_ApJ}. By comparing the deconvolution results with their initial gas composition, we identified newly formed gas products, listed in Table~\ref{table1}. Components with abundances below $1\times 10^{-10}$ are near the noise level and considered negligible. Table~1 therefore summarizes the newly formed inorganic and organic gaseous products with abundance above $1\times 10^{-10}$. 

\begin{table*}[t]
\centering
\caption{Deconvolution results for newly formed gas‑phase species in the six experiments. Note that the reported relative abundances are semi-quantitative estimates, focusing on species identification rather than quantification, and they should be interpreted with caution.}
\label{table1}
\renewcommand{\arraystretch}{1.3}
\newcolumntype{C}{>{\centering\arraybackslash}p{2.1cm}}

\begin{tabular}{|c|c|C|C|C|C|C|C|}
\hline
\multicolumn{2}{|c|}{\multirow{2}{*}{Gas-phase Products}} & \multicolumn{6}{c|}{Relative Abundance} \\ \cline{3-8} 
\multicolumn{2}{|c|}{} & CO$_2$-rich/CO & CO$_2$-rich/CH$_4$ & N$_2$-rich/CO & N$_2$-rich/CH$_4$ & H$_2$O-rich/CO & H$_2$O-rich/CH$_4$ \\ \hline

\multirow{9}{*}{Inorganics} & H$_2$     & $1.60 \times 10^{-10}$ & $3.62 \times 10^{-8}$  & $5.08 \times 10^{-10}$ & $1.63 \times 10^{-7}$ & $1.96 \times 10^{-7}$ & $2.03 \times 10^{-6}$ \\
                            & NH$_3$    & $5.18 \times 10^{-10}$ & $4.62 \times 10^{-9}$  & \textendash            & $1.19 \times 10^{-8}$ & $3.21 \times 10^{-9}$ & $3.49 \times 10^{-8}$ \\
                            & HN$_3$    & \textendash            & $6.60 \times 10^{-10}$ & \textendash            & \textendash           & \textendash           & \textendash           \\
                            & CO        & \textendash            & $6.86 \times 10^{-7}$  & \textendash            & $1.04 \times 10^{-8}$ & \textendash           & $3.00 \times 10^{-10}$ \\
                            & NO        & $8.16 \times 10^{-10}$ & \textendash            & $2.18 \times 10^{-8}$  & $3.71 \times 10^{-8}$ & $7.25 \times 10^{-10}$ & $1.10 \times 10^{-8}$ \\
                            & H$_4$N$_2$& \textendash            & $2.57 \times 10^{-10}$ & \textendash            & \textendash           & \textendash           & \textendash           \\
                            & N$_2$O    & $5.74 \times 10^{-8}$  & $3.05 \times 10^{-8}$  & $5.05 \times 10^{-8}$  & $2.25 \times 10^{-9}$ & $3.02 \times 10^{-8}$ & $1.06 \times 10^{-9}$ \\
                            & NO$_2$    & $5.96 \times 10^{-10}$ & $4.57 \times 10^{-10}$ & $2.59 \times 10^{-10}$ & $1.55 \times 10^{-9}$ & $1.34 \times 10^{-10}$ & $7.12 \times 10^{-9}$ \\
                            & HNO$_2$   & $1.31 \times 10^{-10}$ & $2.78 \times 10^{-10}$ & $2.40 \times 10^{-10}$ & \textendash           & \textendash           & $7.61 \times 10^{-10}$ \\ \hline

\multirow{13}{*}{Organics}  & CH$_4$    & $2.70 \times 10^{-10}$ & \textendash            & \textendash            & \textendash           & $1.29 \times 10^{-10}$ & \textendash           \\
                            & HCN       & $1.39 \times 10^{-8}$  & $1.00 \times 10^{-8}$  & $2.57 \times 10^{-8}$  & $1.46 \times 10^{-9}$ & $1.52 \times 10^{-8}$ & $4.86 \times 10^{-8}$ \\
                            & C$_2$H$_4$& $1.03 \times 10^{-10}$ & $2.28 \times 10^{-10}$ & \textendash            & \textendash           & \textendash           & $1.66 \times 10^{-10}$ \\
                            & CH$_3$OH  & \textendash            & $1.03 \times 10^{-10}$ & \textendash            & \textendash           & \textendash           & \textendash           \\
                            & HCHO      & \textendash            & \textendash            & \textendash            & \textendash           & \textendash           & $1.55 \times 10^{-10}$ \\
                            & C$_2$H$_6$& \textendash            & \textendash            & \textendash            & \textendash           & \textendash           & $1.36 \times 10^{-10}$ \\
                            & CH$_3$CHO & $4.04 \times 10^{-10}$ & \textendash            & $8.00 \times 10^{-10}$ & $2.40 \times 10^{-10}$ & $5.14 \times 10^{-10}$ & $1.07 \times 10^{-10}$ \\
                            & CH$_3$OCH$_3$ & $4.97 \times 10^{-10}$ & $2.55 \times 10^{-10}$ & \textendash        & $1.09 \times 10^{-10}$ & $8.64 \times 10^{-10}$ & $5.84 \times 10^{-9}$ \\
                            & CH$_3$NO  & \textendash            & \textendash            & \textendash            & \textendash           & $1.22 \times 10^{-9}$ & \textendash           \\
                            & HCOOH     & \textendash            & $2.53 \times 10^{-10}$ & \textendash            & \textendash           & \textendash           & \textendash           \\
                            & (CH$_3$)$_2$NH & $4.31 \times 10^{-10}$ & $6.61 \times 10^{-10}$ & \textendash        & \textendash           & $2.21 \times 10^{-10}$ & $1.06 \times 10^{-10}$ \\
                            & (CH$_2$)$_2$O  & \textendash        & \textendash            & \textendash            & \textendash           & $1.16 \times 10^{-10}$ & \textendash           \\
                            & C$_3$H$_4$     & \textendash        & \textendash            & \textendash            & \textendash           & \textendash           & $1.21 \times 10^{-10}$ \\ \hline
\multicolumn{8}{|c|}{\footnotesize \textendash: gas product with abundance below $1 \times 10^{-10}$} \\ \hline
\end{tabular}
\end{table*}
\begin{figure*}[t!]
    \centering
    \resizebox{16cm}{!}
    {\includegraphics{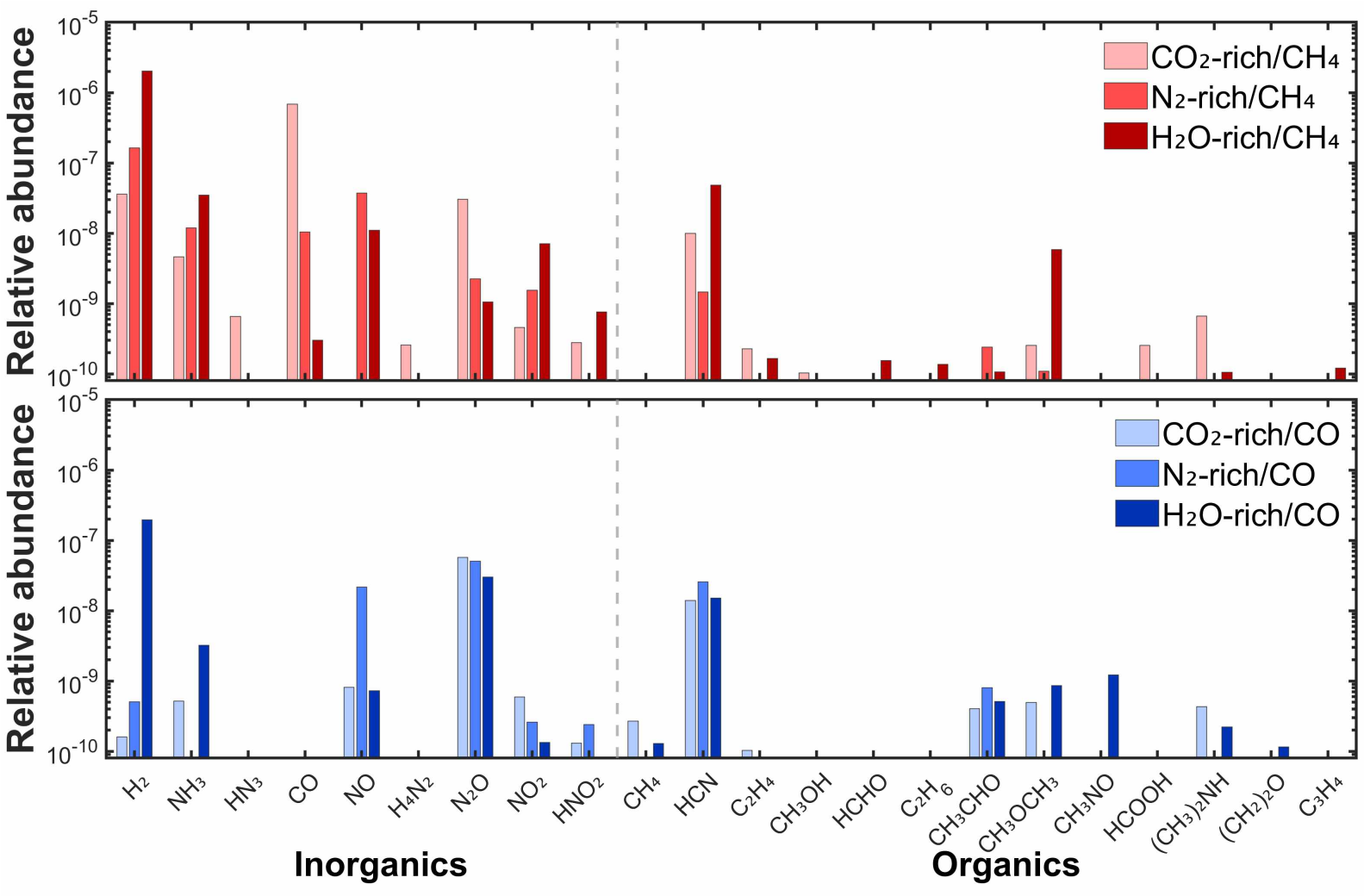}}
    \caption{Comparison of the relative abundances of newly formed gas-phase species in the six experiments. The upper panel shows the three CH$_4$ experiments and the lower panel shows the three CO experiments. Inorganic and organic species are separated by the dashed line.}
    \label{fig3}
\end{figure*}
In the CO$_2$-rich condition with CO as the carbon source, we detected inorganic nitrogen oxides N$_2$O as the dominant products accompanied by oxidized nitrogen species NO, NO$_2$, and HNO$_2$, and reduced nitrogen species such as NH$_3$ and H$_2$. The main organic product was HCN, along with oxygenated hydrocarbons including CH$_3$OCH$_3$ and CH$_3$CHO, the nitrogen-containing compound (CH$_3$)$_2$NH, and the hydrocarbon C$_2$H$_4$. When CH$_4$ was used as the carbon source, CO emerged as the predominant product, with a potential underlying redox reaction like $3\text{CO}_2+\text{CH}_4\rightarrow 4\text{CO}+2\text{H}_2\text{O}$. We also observed strong H$_2$ formation from CH$_4$ and H$_2$O dissociation, with NH$_3$ levels markedly higher, reflecting a shift toward a more reducing environment. Under the CO$_2$-rich condition with the CH$_4$ carbon source, organic products showed greater diversity, including nitrogen- (HCN and (CH$_3$)$_2$NH), oxygen-containing organic species (CH$_3$OCH$_3$, HCOOH, and CH$_3$OH), and hydrocarbon (C$_2$H$_4$).

In the N$_2$-rich/CO experiment, N$_2$O and NO were the predominant inorganic products, while H$_2$, NO$_2$, and HNO$_2$ were also detected. HCN and CH$_3$CHO were identified as the main organic compounds. The N$_2$-rich/CH$_4$ experiment showed significant hydrogen production from methane dissociation, accompanied by various nitrogen oxides (e.g., NO, N$_2$O, NO$_2$). A key difference from the N$_2$-rich/CO experiment was the observation of the reductive nitrogen-containing gas NH$_3$. Among the organic products, HCN was the predominant species, along with small amounts of oxygenated compounds such as CH$_3$CHO and CH$_3$OCH$_3$.

In the H$_2$O-rich/CO experiment, H$_2$ is the most abundant gas product, probably produced from the photolysis of H$_2$O. Organic products were diverse, including HCN, (CH$_3$)$_2$NH, and oxygenated species such as CH$_3$NO, CH$_3$OCH$_3$, CH$_3$CHO, and (CH$_2$)$_2$O, indicating the incorporation of nitrogen and oxygen into carbon-containing species. In the H$_2$O-rich/CH$_4$ experiment, H$_2$ again dominated the inorganic fraction. NH$_3$ was more abundant than that in the H$_2$O-rich/CO experiment, and multiple nitrogen oxides were observed. Organic products included HCN, (CH$_3$)$_2$NH, CH$_3$OCH$_3$, HCHO, CH$_3$CHO, and hydrocarbons (C$_2$H$_4$, C$_2$H$_6$, C$_3$H$_4$). The presence of CH$_4$ promoted abundant hydrocarbon formation and increased nitrogen  and oxygen-containing organics.

Overall, for the three groups of experiments, the CH$_4$ cases always produced a wider variety of inorganic and organic products compared to the CO cases, including higher abundance of reduced nitrogen (NH$_3$) and hydrocarbons. This trend is consistent with the higher reactivity of CH$_4$ compared to CO, resulting in richer chemical reactions in the gas mixtures. HCN is widely considered an important precursor for haze formation. However, our results indicate that haze production rate is not directly correlated with HCN abundance alone. For example, HCN is the least abundant in the N$_2$-rich/CH$_4$ experiment among the six experiments, but its haze production rate is the second highest, as shown in Section 3.2. This suggests that HCN is likely one of several contributing precursors, and the overall distribution of gas-phase products must be considered.

\subsection{Density, Production Rate, and Size Distribution of Haze Particles}

After the experiments concluded, only the three CH$_4$ experiments produced sufficient solid samples for collection and mass measurement. Among them, only the N$_2$-rich/CH$_4$ and H$_2$O-rich/CH$_4$ cases produced enough materials for density analysis. The determined particle densities for the N$_2$-rich/CH$_4$ and H$_2$O-rich/CH$_4$ samples are 1.50 g cm$^{-3}$ and 1.51 g cm$^{-3}$, respectively, with uncertainties below 0.3\%, based on 20 replicate measurements. The densities of these two haze analogs fall at the upper end of the range reported for haze particles produced under different planetary atmospheric conditions using the same experimental setup \citep{2017_He_ApJL, 2024_He_NatAs, 2025_Wang_ApJ}. Because particle density is strongly controlled by chemical composition, these differences indicate distinct compositional characteristics among hazes formed under varying atmospheric regimes. Even with an identical setup, variations in the initial gas mixture and temperature can significantly alter haze composition, as demonstrated in previous work \citep{2020_Moran_PSJ}. Previous studies have shown that haze particles contain thousands of organic compounds, and that their density is influenced by factors such as elemental composition, molecular size, polarity, and degree of unsaturation. The relatively high densities observed in this study suggest that these haze analogs are composed of larger and more polar molecules. These characteristics will be examined in detail in the following sections, further demonstrating that variations in atmospheric composition produce haze particles with distinct chemical properties.

The atmospheric compositions also strongly influence the haze production rate. For the three experiments with CH$_4$, production rates determined from the collected particle masses are 0.07 mg h$^{-1}$, 1.44 mg h$^{-1}$, and 1.49 mg h$^{-1}$ for the CO$_2$-rich/CH$_4$, N$_2$-rich/CH$_4$, and H$_2$O-rich/CH$_4$ cases, respectively. The markedly lower production rate for the CO$_2$-rich/CH$_4$ experiment is likely due to its more oxidizing atmosphere, where CH$_4$ preferentially reacts with abundant CO$_2$ to form volatile gaseous products rather than the solid organic hazes. This interpretation is supported by the significantly elevated CO levels observed in the gas phase via RGA measurements, as shown in Fig.~\ref{fig3}. For the three experiments with CO, the haze production rates are too low to generate weighable mass. The haze particles deposited on the mica substrates were examined by AFM for estimating the production rates.

\begin{figure}[ht!]
   \centering
   \includegraphics[width=\hsize]{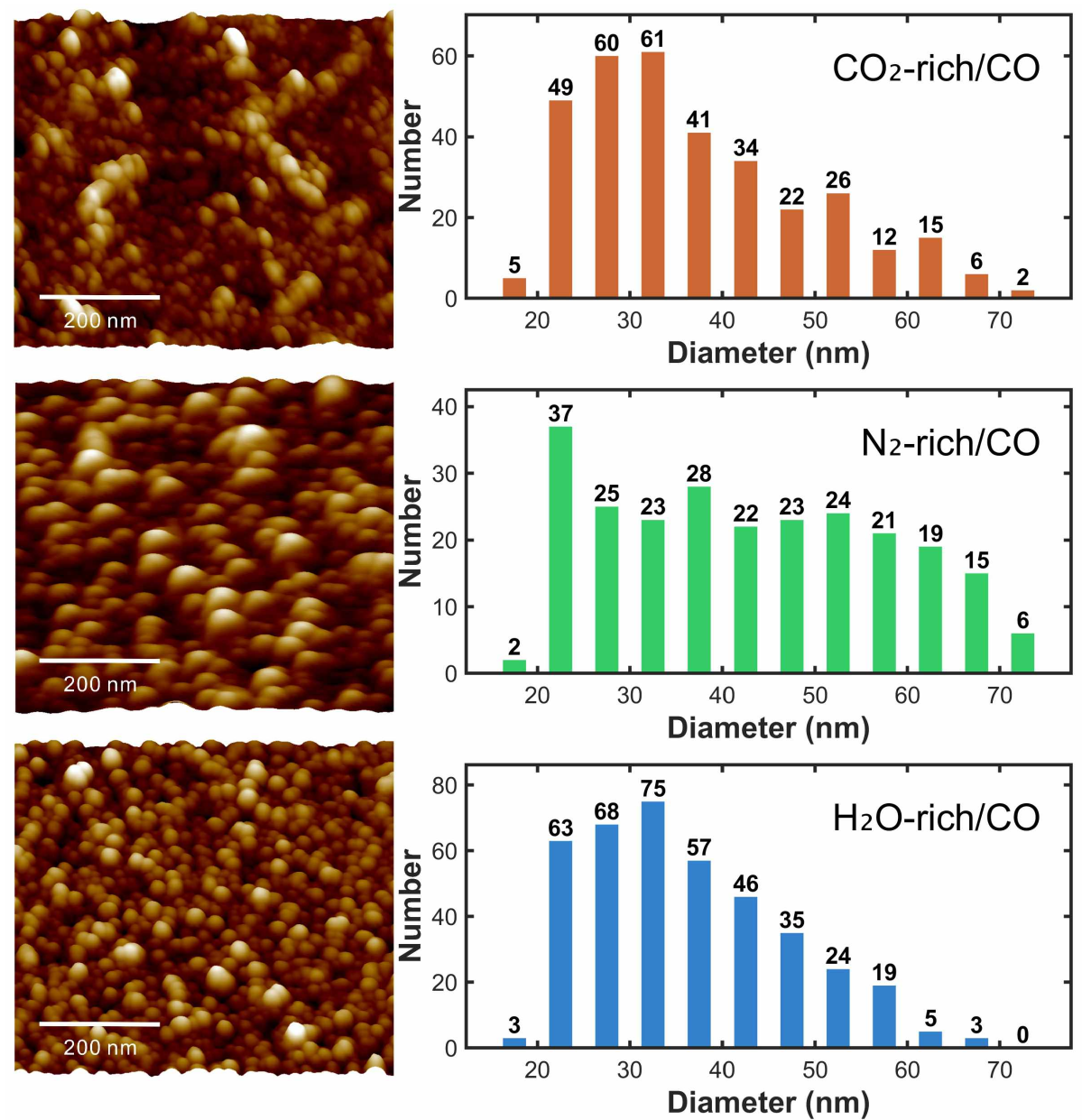}
   \caption{Morphology and particle size distributions of haze particles formed in the three CO experiments. The left panels show AFM images of three film samples produced in the CO$_2$-rich/CO (upper), N$_2$-rich/CO (middle), and H$_2$O-rich/CO (lower) experiments. The right panels present the particle size and number distributions derived from the corresponding images.}
   \label{fig4}
\end{figure}

As shown in Fig.~\ref{fig4}, AFM images of the three film samples indicate that each experiment produced at least one layer of haze particles, and these particles are approximately spherical. However, both the particle size and abundance differ significantly across the three experimental conditions. In the CO$_2$-rich/CO experiment, the particles have an average diameter of 37.06 nm, with more than 95\% falling within the 20-70 nm range. Only a few particles exceeded 70 nm in diameter, likely formed by aggregation of smaller monomers. In the N$_2$-rich/CO experiment, the average diameter increases to 42.92 nm, with a more uniform size distribution in the range of 20-70 nm. Particles in the H$_2$O-rich/CO experiment have an average diameter of 36.07 nm and are mainly concentrated in the 20-60 nm range. Overall, the haze particles produced in the CO experiments are predominantly distributed within the 20-70 nm range.

\begin{figure}[ht!]
   \centering
   \includegraphics[width=\hsize]{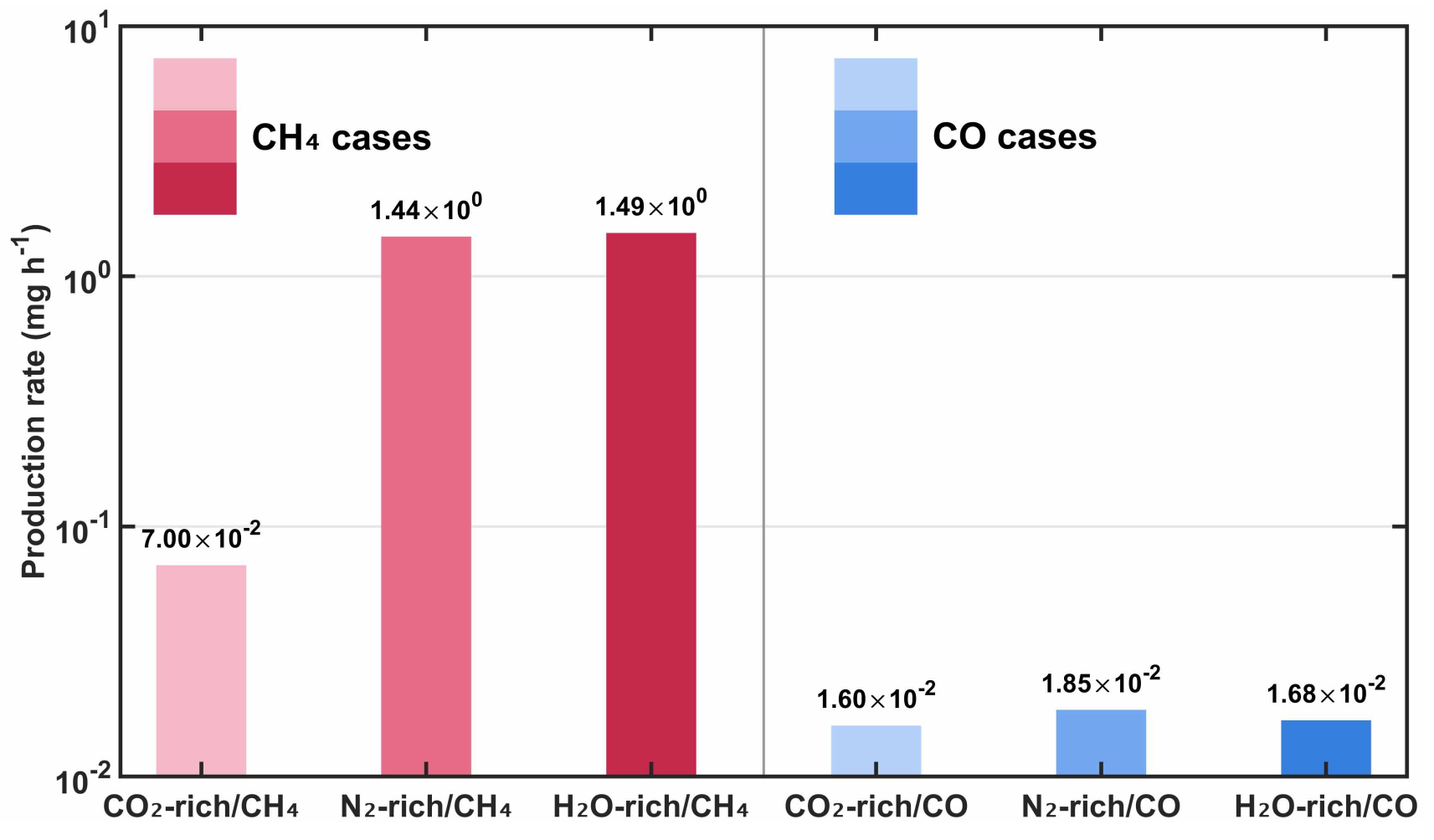}
   \caption{Haze production rates for six experiments. The red bars (left panel) and blue bars (right panel) correspond to experiments using CH$_4$ and CO as the carbon source, respectively. Within each panel, the light-to-dark color gradient denotes the dominant background gas: CO$_2$-rich (lightest), N$_2$-rich (intermediate), and H$_2$O-rich (darkest).}
   \label{fig5}
\end{figure}

To evaluate the haze production rates, we estimated the mass of collected haze particles based on the particle size distribution obtained from the AFM analysis and derived the corresponding lower limits of the haze production rates in the three CO experiments. The calculated haze production rates are 1.60$\times$10$^{-2}$, 1.85$\times$10$^{-2}$, and 1.68$\times$10$^{-2}$ mg h$^{-1}$ for the CO$_2$-rich/CO, N$_2$-rich/CO, and H$_2$O-rich/CO experiments, respectively. As shown in Fig.~\ref{fig5}, these production rates are substantially lower than those observed in the corresponding CH$_4$ experiments, highlighting the markedly different efficiencies of CO and CH$_4$ as carbon sources in driving organic haze formation. The different efficiency of the CO- and CH$_4$-driven photochemistry also suggests that distinct reaction pathways govern particle nucleation and growth, yielding fundamentally different chemical compositions in the resulting haze particles. Therefore, to elucidate the molecular mechanisms underlying these discrepancies, we next examine the chemical composition of the produced haze particles.

\subsection{ FTIR Spectra of Haze Particles}

\begin{figure}[ht!]
   \centering
   \includegraphics[width=\hsize]{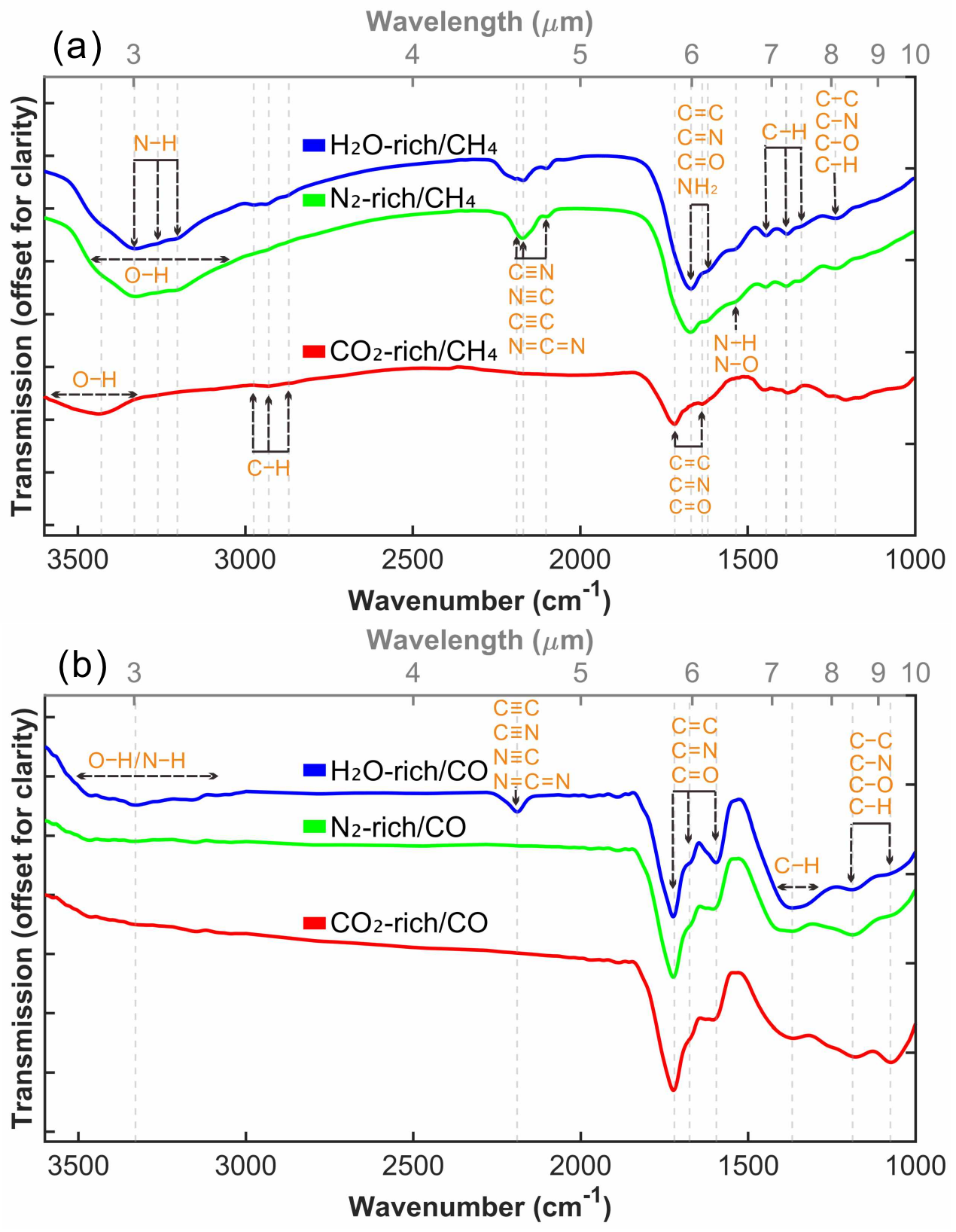}
   \caption{Infrared transmittance spectra of haze samples from the six experiments. (a) Transmittance spectra (3600--1000~cm$^{-1}$) of haze particles formed with CH$_4$ as the carbon source. (b) Transmittance spectra (3600--1000~cm$^{-1}$) of haze particles formed with CO as the carbon source. The blue, green, and red lines represent the H$_2$O-rich, N$_2$-rich, and CO$_2$-rich conditions, respectively. Spectra are vertically offset and rescaled for clarity, with all peaks within each spectrum multiplied by the same constant factor so that relative peak intensities and peak positions are preserved.}
   \label{fig6}
\end{figure}

Fig.~\ref{fig6} shows the transmittance spectra of the haze particles produced in the six simulated atmospheres, revealing a rich diversity of functional groups. The haze particles from the H$_2$O-rich/CH$_4$ and N$_2$-rich/CH$_4$ experiments exhibited similar absorption features, suggesting that CH$_4$-driven photochemistry proceeds in a comparable manner in both background atmospheres. Multiple functional groups were identified, including O--H, N--H, C--H, C$\equiv$N, N$\equiv$C, C$\equiv$C, N=C=N, C=C, C=N, C=O, N--O, C--C, C--N, and C--O \citep{1991_LinVien_HandbookIRRaman}. The functional groups identified in these samples, along with their absorption characteristics, closely resemble those reported for haze particles in simulated water-rich atmospheres by previous studies \citep{2024_He_NatAs}, likely reflecting comparable initial gas compositions and photochemical pathways. In contrast, the CO$_2$-rich/CH$_4$ sample exhibits distinct absorption features, generally with lower intensities. Notably, absorptions corresponding to triple-bond species such as C$\equiv$N, N$\equiv$C, and C$\equiv$C are nearly absent, reflecting a reduced formation of these highly unsaturated functional groups under CO$_2$-rich conditions. Additionally, the absorption bands associated with O--H, C=C, C=N, C=O, and N--H were blue-shifted relative to the H$_2$O-rich/CH$_4$ and N$_2$-rich/CH$_4$ cases. Such shifts indicate increased vibrational frequencies associated with larger bond force constants, which may result from the presence of adjacent electron-withdrawing groups and/or the formation of more rigid and highly cross-linked molecular structures. The systematic blueshift indicates that haze particles formed in the CO$_2$-rich/CH$_4$ experiments possess more oxidized and structurally constrained organic networks, consistent with the more oxidizing nature of the initial gas mixture. These results highlight the strong dependence of haze chemical composition on atmospheric oxidation state.

When CO was used as the carbon source, the overall oxidation state of the initial gas composition increases compared to the CH$_4$ cases, and therefore the resulting haze compositions are expected to differ. As shown in Fig.~\ref{fig6}, the three CO-derived samples display almost no detectable C--H stretching absorptions in the 3000--2800 cm$^{-1}$ region, indicating that oxidizing conditions strongly suppress the formation of hydrocarbon species. A striking feature in the infrared spectra is that absorption in the double-bond stretching region (the vibrational modes of C=O, C=N, and C=C at \textasciitilde 1600--1700 cm$^{-1}$) becomes dominant in all three CO-derived samples, as well as in the CO$_2$-rich/CH$_4$ sample, suggesting a lower degree of saturation in hazes formed under relatively oxidizing conditions. The CO$_2$-rich/CO and N$_2$-rich/CO hazes exhibit highly similar spectral characteristics,  with very weak O--H and N--H stretching absorptions, while bands in the double-bond region and the fingerprint region remain prominent. This pattern suggests that haze particles generated in these highly oxidizing atmospheres contain relatively unsaturated yet oxygen- and nitrogen-containing networks, with limited incorporation of hydrogen-bearing groups. In contrast, the H$_2$O-rich/CO sample displays additional spectral features, including pronounced O--H (3500--3000 cm$^{-1}$), N--H (3300--3200 cm$^{-1}$), and triple-bond (2200--2100 cm$^{-1}$) absorptions. The presence of these functional groups indicates that the relatively less oxidizing H$_2$O-rich composition allows chemical pathways to preserve or generate more hydrogen-bearing and highly unsaturated chemical structures. These differences reinforce the conclusion that the chemical evolution of photochemical hazes is strongly governed by the oxidation state of the atmosphere; progressively more oxidizing environments---CO$_2$-rich > N$_2$-rich > H$_2$O-rich---yield haze particles with lower hydrogen content and constrained functional-group diversity.

Overall, comparison between the CH$_4$- and CO-derived hazes reveals systematic differences in infrared spectral features and inferred molecular structures, reflecting the influence of the initial gas composition. The more reducing initial gas mixtures in the CH$_4$ experiments yields hazes with a greater diversity of hydrogen-rich functional groups, consistent with their higher haze production rate and the broader range of gas products detected by the RGA measurements. These observations collectively highlight the strong link between atmospheric redox state, photochemical pathways, and the resulting chemical composition of the haze particles. To further resolve the molecular-level complexity and chemical makeup of these hazes, we next turn to VHRMS analyses, which provide complementary insights into their molecular formulas, heteroatom incorporation, and the distribution of unsaturation.

\subsection{VHRMS Analysis of Haze Particles}

\begin{figure*}[t!]
    \centering
    \resizebox{16cm}{!}
    {\includegraphics{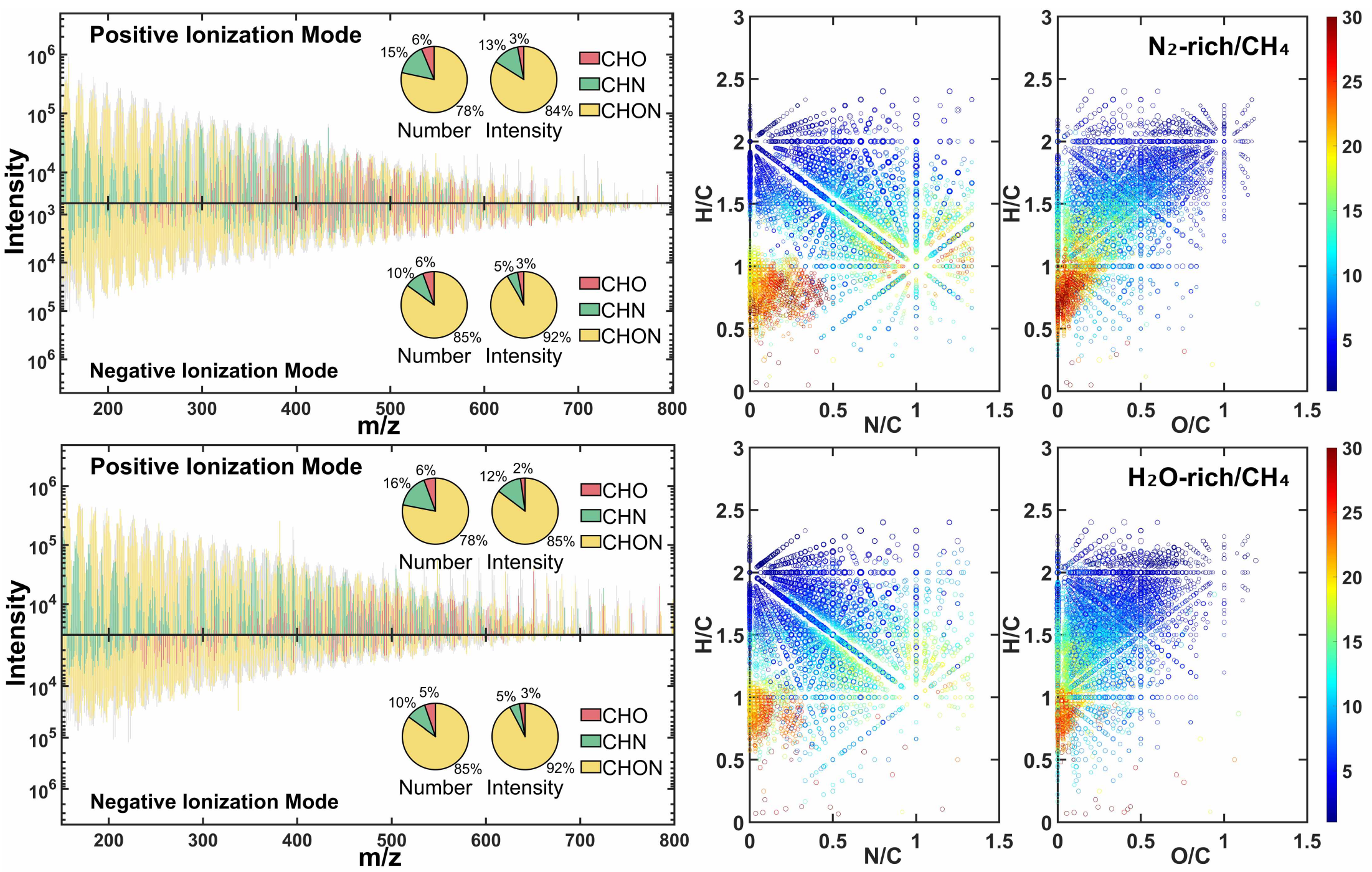}}
    \caption{Molecular composition of haze samples from the N$_2$-rich/CH$_4$ and H$_2$O-rich/CH$_4$ experiments. Left panels: Very high-resolution mass spectra (m/z 150--800) of haze particles produced from the N$_2$-rich/CH$_4$ and H$_2$O-rich/CH$_4$ experiments, acquired in both positive and negative ionization modes. Assigned CHO, CHN, and CHON species are shown in red, green, and yellow, respectively, while grey data points correspond to mass peaks with no formula matched within the mass accuracy threshold ($<$~0.0001~Da). For each spectrum, the adjacent pie charts summarize the percentage of molecular formulas and their relative abundances based on summed peak intensities of the assigned organic species. Right panels: Corresponding van Krevelen diagrams combining matched formulas from both ionization modes, shown as H/C vs. N/C and H/C vs. O/C. Each circle represents an assigned formula, where color denotes the degree of unsaturation and marker size scales with peak intensity.}
    \label{fig7}
\end{figure*}

Fig.~\ref{fig7} shows the results from VHRMS analysis for the N$_2$-rich/CH$_4$ and H$_2$O-rich/CH$_4$ haze samples. We focused on these two cases with high production rates, because the other cases resulted in mass signals at or below the noise level due to their low haze production rates and limited haze solubility in methanol. We obtained the mass spectra in both positive and negative ionization modes to cover more comprehensive species as some molecules can be only detected in positive or negative mode depending on their acid-base properties. As illustrated in the left panels of Fig.~\ref{fig7}, both haze samples exhibit thousands of distinct peaks across both modes, revealing their broad molecular diversity and chemical complexity. Based on the precise masses, we assigned molecular formulas to these peaks into three subgroups: CHO, CHN, and CHON, which are shown with different colors in Fig.~\ref{fig7}. Specifically, for the N$_2$-rich/CH$_4$ solid sample, we identified 2518 and 4252 molecular formulas in positive and negative modes, respectively, whereas for the H$_2$O-rich/CH$_4$ sample, 3476 and 2565 formulas were detected in the corresponding modes. Across both samples, the mass spectra show a clear approximately 14 Da repeating pattern, consistent with CH$_2$ spacing. Such periodicity is characteristic of homologous organic series and has been observed for similar organic haze samples \citep[e.g.,][]{2017_Gautier_P&SS, 2020_Jovanovic_Icar, 2022_Moran_JGRE}, reflecting molecular growth through repeated monomer addition and substitution reactions.

Further analysis of these identified formulas in the three subgroups revealed that the samples were dominated by the CHON subgroup, suggesting that nitrogen and oxygen are preferentially co-incorporated into molecules forming hazes. In contrast, the CHN and CHO subgroups constituted much smaller fractions. To capture the bulk chemical properties of these complex mixtures, we calculated the intensity-weighted average molecular formula, which are C$_{12.4}$H$_{18.3}$O$_{3.0}$N$_{5.8}$ with a molecular weight of 296.4 Da, and C$_{13.6}$H$_{21.4}$O$_{2.8}$N$_{5.4}$ with a molecular weight of 305.2 Da for the N$_2$-rich/CH$_4$ and H$_2$O-rich/CH$_4$ samples, respectively. The H$_2$O-rich/CH$_4$ sample displayed a slightly larger molecular size but lower O/C (0.21 vs. 0.24) and N/C (0.40 vs. 0.47) ratios, indicating shifts in chemical pathways driven by the different initial gas compositions.

To further visualize the elemental composition distributions, we applied Van Krevelen diagrams (H/C vs. N/C and O/C plots), as shown in the right panels of Fig.~\ref{fig7} \citep{2009_VanKrevelen_PropertiesOfPolymers}. The identified molecular formulas in both samples spanned a wide distribution of N/C ratios, while their O/C ratios were generally below 1. Several focal points were evident in the diagrams at (N/C, H/C) coordinates of (1, 1) and (0, 2), and at (O/C, H/C) coordinates of (0, 1), (1, 2), and (0, 2). Molecules radiate outwards from these coordinates, indicating that simple precursors, such as HCN, CH$_2$O, and C$_2$H$_4$, act as building blocks for complex haze chemistry. Similar radiating patterns were also observed in the mass spectra of organic hazes in previous studies \citep[e.g.,][]{2022_Moran_JGRE, 2025_Wang_ApJ}. As demonstrated by color bar in the van Krevelen diagrams, molecular unsaturation degree varies widely from 0 to 30 for both samples, reflecting complex chemical structures. However, their distributions differ at equivalent unsaturation levels. The N$_2$-rich/CH$_4$ sample is enriched in highly unsaturated molecules (H/C $<$ 1, N/C $<$ 0.5, O/C $<$ 0.5), whereas the H$_2$O-rich/CH$_4$ sample contains more relatively saturated species (H/C $>$ 1.5). These contrasting patterns collectively suggest a greater prevalence of double or triple bonds in the N$_2$-rich sample, consistent with its lower average H/C ratio (1.48 vs. 1.57). Overall, the VHRMS analysis reveals distinct molecular fingerprints for the two samples, demonstrating that the initial atmospheric composition plays a primary role in shaping the chemical pathways and the resulting haze molecular complexity.

\subsection{Discussion}

In planetary atmospheres, CH$_4$ is believed to be key for haze formation. However, recent laboratory studies have demonstrated that organic haze particles were produced in simulated atmospheres without CH$_4$, in which carbon monoxide served as an alternative carbon source \citep{2018_He_AJ, 2018_Horst_NatAs, 2020_He_PSJ, 2020_Moran_PSJ}. Due to the compositional complexity of the initial gas mixtures used in earlier studies, it was not possible to directly compare the efficiencies of CH$_4$ and CO as carbon sources in driving organic haze formation or to elucidate their distinct chemical pathways. Here we conducted experiments under the identical conditions and background gases to disentangle the role of CH$_4$ and CO. Our results demonstrate that the haze production rate in the CH$_4$ experiments is much higher than that in the corresponding CO experiments. The N$_2$-rich/CH$_4$ and H$_2$O-rich/CH$_4$ experiments produce haze approximately 80 and 90 times more efficiently than the respective CO cases. Even in the more oxidizing CO$_2$-rich/CH$_4$ gas mixture, the haze product rate is about five times higher than the CO case. Irrespective of the background gas composition, CH$_4$ consistently proves to be a significantly more efficient carbon source than CO. This observation can be attributed to distinct reaction pathways in differing redox state of the simulated atmospheres. The more reducing atmospheres with CH$_4$ produce more diverse gas precursors (Table~\ref{table1}), which could polymerize and incorporate into N/O-containing organic haze, as evidenced by enhanced C--H, N--H, and O--H absorptions in the FTIR spectra of the CH$_4$-derived haze samples. In contrast, the CO experiments preferentially generate highly oxidizing O-bearing species, resulting in significantly lower organic haze yields compared to the reducing CH$_4$ cases.

Interestingly, our H$_2$O-rich/CH$_4$ experiment has the highest haze production rate. Enhanced haze production under H$_2$O-rich conditions has also been reported previously by \citep{2018_Horst_NatAs}. More broadly, classic prebiotic synthesis experiments have long shown that H$_2$O can support the formation of organic compounds, as demonstrated in the Miller-Urey experiment \citep{1953_Miller_Sci} and in later photochemical experiments performed in the presence of water \citep{2022_Zang_AsBio}. The specific mechanism responsible for the enhanced haze yield in our H$_2$O-rich/CH$_4$ experiment remains uncertain, but one possible interpretation is that it reflects a competition between the oxidizing effect of OH radicals and the reducing effect of free H generated during H$_2$O photolysis. In particular, the photochemical model of Pinto et al. identified H atoms as key agents for forming organic precursors formaldehyde, which could in turn promote haze formation \citep{1980_Pinto_Sci}. In this framework, H released from water photolysis can promote the formation of oxygenated organic precursors that may subsequently participate in further growth toward more complex organic material. At the same time, OH radicals are expected to compete with this process through oxidation pathways. Regarding the role of hydrogen in haze formation, previous experiments showed that adding abundant H$_2$ reduced haze production, and they proposed that H$_2$ saturates hydrocarbon chains and prevents molecules from becoming large enough to condense \citep{2009_DeWitt_AsBio}. However, the molecular H$_2$ in that previous study may act differently from the highly reactive H produced by H$_2$O photolysis in our experiment. In our H$_2$O-rich experiment, water photolysis likely introduces free H that can participate directly in precursor formation rather than merely terminating growth, thereby maintaining a more reducing radical environment that is more favorable to organic growth. Consistent with this interpretation, our VHRMS results show that the H/C ratio of the H$_2$O-rich/CH$_4$ haze sample (1.57) is higher than that of the N$_2$-rich/CH$_4$ sample (1.48), suggesting that hydrogen derived from water is incorporated into the haze chemistry without suppressing particle growth. Our results suggest that, under the reducing conditions of the H$_2$O-rich/CH$_4$ experiment, H-driven precursor formation may outpace OH-driven oxidation, allowing a net buildup of haze precursors and ultimately higher particle production. This interpretation remains tentative and will require detailed mechanistic investigation in future work.

The production rate determines the initial supply of haze particles to an atmosphere, and it is a key input in exoplanet modeling, where it is often parameterized over several orders of magnitude as a source term in the upper atmosphere (e.g., \citealt{2016_Arney_Astro}; \citealt{2019_Lavvas_APJ}; \citealt{2023_Gao_ApJ}). Although the haze production rates derived from our laboratory experiments cannot be directly used in atmospheric models given the differences in conditions, they nevertheless provide valuable constraints on the relative haze production efficiencies across different gas compositions. Importantly, haze production rate alone does not determine the atmospheric distribution of haze particles, which also depends on subsequent microphysical evolution. Once formed, haze particles undergo growth, coagulation, transport, and sedimentation, and these microphysical pathways are strongly modulated by their intrinsic physical properties \citep{2021_Gao_JGRE}. Among these, density and particle size play central roles in governing haze evolution. Our measurements show that the N$_2$-rich/CH$_4$ and H$_2$O-rich/CH$_4$ haze samples exhibit higher densities than both typical Titan tholins and previously studied water-rich exoplanet haze materials \citep{2017_He_ApJL, 2024_He_NatAs}. Higher-density particles experience stronger gravitational settling forces, which would increase their settling velocities and favor their accumulation in deeper atmospheric layers. 

However, the impact of density is coupled to particle size, which evolves dynamically through aggregation, condensation, and coagulation. Our AFM measurements indicate that the CO-derived haze particles range from 10--80 nm, consistent with the monomer sizes for exoplanet hazes reported in prior work \citep{2018_He_ApJL, 2018_He_AJ, 2020_He_NatAs, 2020_He_PSJ}. This size distribution provides insight into early-stage microphysical processes governing nucleation, aggregation, and subsequent growth. This size range is also similar to that reported by \citet{2014_Horst_ApJ}, who found that adding CO to simulated atmospheres increased both particle size and number density. Moreover, particles in this range fall within the Rayleigh scattering regime, leading to a strong wavelength dependence in opacity, especially towards short optical and UV wavelengths. This scattering behavior has direct implications for the interpretation of both transmission and reflection spectra of exoplanet atmospheres. For such small haze particles, scattering is expected to dominate at shorter wavelengths, whereas in the infrared, including much of the JWST range, the spectral effect is expected to be controlled primarily by particle absorption. Therefore, different haze compositions could be distinguished through their infrared absorption features in future JWST observations \citep{2024_He_NatAs, 2025_Jaziri_A&A}. Together, the measured density and particle size establish realistic boundary conditions for microphysical and radiative transfer modeling, and are crucial for interpreting observations of haze-bearing terrestrial exoplanet atmospheres.

Building on the physical constraints discussed above, the chemical composition of the haze particles provides an additional layer of insight into how atmospheric oxidation state shapes haze formation and evolution. FTIR analysis shows that hazes generated from CH$_4$-containing atmospheres exhibit higher functional-group diversity and structural complexity. Among the three CH$_4$ experiments, the N$_2$-rich and H$_2$O-rich cases display broadly similar spectral features, indicating that CH$_4$-driven photochemistry follows comparable pathways despite differences in background gas composition. In contrast, hazes produced in the CO$_2$-rich atmosphere display simpler chemical signatures. This reduced complexity with increasing CO$_2$ abundance is consistent with a previous study showing that organic products formed in N$_2$/CH$_4$/CO$_2$ mixtures became more oxidized and less chemically complex as the CO$_2$ fraction increased \citep{2004_Trainer_AsBio}. Such behavior is expected in a relatively oxidizing chemical environment, where gas-phase redox reactions may suppress the formation and polymerization of larger organic molecules. The three CO experiments similarly produce haze particles with lower yields and limited functional-group diversity, further demonstrating that atmospheric redox state plays a critical role in photochemical pathways and haze composition.

These compositional trends have direct implications for exoplanet characterization. Highly oxidized atmospheres (e.g., CO$_2$-dominated or CH$_4$/H$_2$-poor) are expected to host sparse and optically weak hazes, whereas reducing, CH$_4$-containing atmospheres favor the formation of optically thick hazes. FTIR spectra show that haze chemical composition varies with initial gas composition, displaying distinct spectral features. These compositional and spectral differences influence transmission, emission, and reflected-light spectra of terrestrial exoplanets. Previous studies have shown that hazes formed under varying atmospheric compositions exhibit substantial differences in their optical constants, resulting in markedly different spectral impacts \citep{2017_Gavilan_APJL, 2023_Corrales_APJL, 2024_Drant_AA, 2025_Li_APJL}. \citet{2026_Drant_A&A} found no significant variations in optical constants when varying the CO/CH$_4$ ratio in the gas phase, whereas \citet{2024_Drant_AA} observed substantial changes upon the addition of CO$_2$. Our compositional results provide a chemical perspective on these observations. In our experiments, changing the carbon source from CH$_4$ to CO affects haze chemistry at the molecular level, including the degree of saturation and the distribution of functional groups. Future optical property study is required to determine whether such compositional changes will translate into large differences in bulk optical properties. By contrast, our CO$_2$-rich experiments, like those of \citet{2004_Trainer_AsBio}, produce chemically simpler and more oxidized products, which likely have distinct spectral signatures as demonstrated by \citet{2024_Drant_AA}. To accurately account for these effects in modeling and observation analysis, future work will focus on measuring the optical constants of our representative haze analogs across observable wavelengths. Such measurements will enable more realistic radiative transfer simulations and strengthen the interpretation of upcoming observations from JWST, ELTs, and future direct-imaging missions---bridging laboratory results to exoplanet atmospheric characterization.

It is worth noting that our experimental framework is designed primarily to compare the roles of CH$_4$ and CO as carbon sources for haze formation, with CO$_2$ treated largely as a background oxidant. Yet CO$_2$ itself can participate directly in organic synthesis, as demonstrated by \citet{2017_Fleury_E&PSL}, who observed organic growth in N$_2$/CO$_2$/H$_2$ mixtures without added CH$_4$ or CO. Additional support comes from \citet{2026_Sohier_A&A}, who showed in H$_2$-dominated experiments that CH$_4$ favors hydrocarbon growth, whereas CO and CO$_2$ also contribute carbon while shifting the chemistry toward more oxidized organic products. \citet{2026_Christensen_ESC} further showed that varying CO$_2$ abundance can lead to large changes in the overall molecular composition of organic hazes. These results highlight that CO$_2$ is not merely a passive oxidant but can also influence haze chemistry. Disentangling the relative contributions of these different carbon sources to haze composition across the diverse atmospheric environments of terrestrial exoplanets remains an important objective for future work.

Finally, VHRMS results highlight the chemical complexity of the N$_2$-rich/CH$_4$ and H$_2$O-rich/CH$_4$ haze samples, reinforcing the trends inferred from the FTIR and RGA measurements. Van Krevelen diagrams reveal chemical pathways involving HCN, C$_2$H$_4$, and CH$_2$O, which are detected by the RGA in the gas phase. These gas species likely act as early intermediates, undergoing polymerization and condensation reactions that drive the formation of progressively larger and more complex molecules as identified in VHRMS. Notably, HCN and CH$_2$O are also important precursors for producing prebiotic biomolecules, such as amino acids, nucleobases, and sugars \citep{1984_Schwartz_OrLi, 2008_Cleaves_PreR}. Thus, we compared the detected molecular formulas to known biomolecules and identified formulas consistent with four biological amino acids and one nucleobase---histidine (C$_6$H$_9$O$_2$N$_3$), tyrosine (C$_{9}$H$_{11}$O$_3$N), arginine (C$_6$H$_{14}$O$_2$N$_4$), tryptophan (C$_{11}$H$_{12}$O$_2$N$_2$), and guanine (C$_5$H$_5$ON$_5$). We also found formulas consistent with several amino acid derivatives, nucleobase derivatives, and non-proteinogenic amino acids. These findings underscore that atmospheric chemistry in reducing planetary environments may be capable of producing fundamental building blocks for life \citep{2012_Horst_AsBio, 2024_Pearce_PSJ}. Future analyses using gas/liquid chromatography mass spectrometry (GC-MS or LC-MS) will enable structural confirmation and quantitative assessment of these compounds.

\section{Conclusion}

In this work, we performed a series of laboratory experiments to investigate haze formation in atmospheres of various terrestrial exoplanets, including H$_2$O-rich, N$_2$-rich, and CO$_2$-rich environments with either CH$_4$ or CO as the carbon source. Our results show that the initial atmospheric redox state strongly regulates haze production rate, physical properties, and chemical compositions. Across all conditions, CH$_4$ consistently acts as a far more efficient carbon source than CO, generating higher haze yields, richer gas-phase chemistry, and greater molecular diversity. N$_2$-rich/CH$_4$ and H$_2$O-rich/CH$_4$ conditions produce haze at rates 80--90 times higher than their CO counterparts, while even the more oxidizing CO$_2$-rich/CH$_4$ experiment produces fivefold greater yields. The resulting haze particles have relatively smaller size and higher density, reflecting their distinct growth pathways and chemical compositions. The CH$_4$-derived hazes contain complex molecular structures, supported by broad FTIR functional group features and thousands of formulas with homologous series patterns revealed by VHRMS. In contrast, CO-derived hazes are chemically simpler, consistent with more oxidizing conditions suppressing the growth of organic molecules.

These findings provide experimentally grounded constraints for haze microphysics and radiative transfer models, enabling more accurate interpretation of exoplanet spectra from current and future missions such as JWST, ELTs, and direct-imaging observatories. Future study will expand on these results by measuring the optical constants of the produced hazes and incorporating them into models to quantify their spectral impacts. Additionally, the detection of several molecular formulas consistent with prebiotic species suggests that reducing atmospheres may support organic chemistry relevant to the origin of life on terrestrial exoplanets, although structural confirmation of these species will require further analysis.
\begin{acknowledgements}
The authors gratefully acknowledge the support from the National Natural Science Foundation of China (42475132). S.E.M. is supported by NASA through NASA Hubble Fellowship grant HST-HF2-51563 awarded by the Space Telescope Science Institute, which is operated by the Association of Universities for Research in Astronomy, Inc., for NASA, under contract NAS5-26555. V.V. acknowledges support from the French National Research Agency in the framework of the ``Investissements d'Avenir'' program (ANR-15-IDEX-02), through the funding of the Origin of Life project of the Université Grenoble Alpes and the French Space Agency (CNES) under their ``Exobiologie, Exoplanètes et Protection Planétaire'' program.
\end{acknowledgements}

%

\bibliographystyle{aa}
\bibliography{bibtex/ref}

\end{document}